\documentclass{icrc}

\usepackage{times}
\usepackage{mathptm}
\usepackage{graphicx} 

\newcommand{\sstt}      {$\sin^2 2\theta$}
\newcommand{\dms}       {$\Delta m^2$}

\newcommand{\chisq}     {$\chi^2$~}
\newcommand{\dmsq}      {$\Delta m^2$}

\begin{document}

\title{Discriminating between $\nu_\mu\leftrightarrow\nu_\tau$ and
  $\nu_\mu\leftrightarrow\nu_{sterile}$ in atmospheric $\nu_\mu$
  oscillations with the Super--Kamiokande detector.}  

\author[1]{A.~Habig}
\affil[1]{University of Minnesota Duluth}

\correspondence{Alec Habig $<$ahabig@umn.edu$>$}

\firstpage{1}
\pubyear{2001}


\maketitle

\begin{abstract}
  A strong body of evidence now exists for atmospheric $\nu_\mu$
  disappearance oscillations.  Such disappearance could be explained by
  oscillations to either $\nu_\tau$ or a ``sterile'' neutrino ($\nu_s$).
  Super--Kamiokande uses three different methods to distinguish between
  these two scenarios.  First, matter effects would suppress the
  $\nu_\mu\leftrightarrow\nu_s$ oscillation amplitude at high energy.
  Second, oscillation to $\nu_s$ would reduce the overall
  neutral-current neutrino interaction rate.  Third, the smoking gun of
  $\nu_\mu\leftrightarrow\nu_\tau$ oscillations would be the observation
  of $\tau$ appearance resulting from charged-current $\nu_\tau$
  interactions.  The results of these three techniques are presented,
  which strongly favor $\nu_\mu\leftrightarrow\nu_\tau$ oscillations
  over $\nu_\mu\leftrightarrow\nu_s$.
\end{abstract}

\section{Introduction}

The Super--Kamiokande (Super--K) experiment has reported evidence for the
oscillation of muon neutrinos produced in cosmic-ray induced showers in
the atmosphere, via the observation of $\nu_\mu$ disappearance as a
function of neutrino pathlength and energy \citep{evidence,SKupmu}.  The
observed $\nu_e$ signal shows no excess, so is not consistent with
substantial $\nu_\mu\leftrightarrow\nu_e$ oscillation; therefore the
missing $\nu_\mu$ must be changing to an ``invisible'' flavor.  Since a
$\nu_\tau$ charged-current (CC) interaction would produce a $\tau$
lepton, $\nu_\tau$ below the 3.4~GeV $\tau$ production threshold do not
interact via the CC channel.  $\nu_\tau$ above this threshold would
produce $\tau$ leptons, but $\tau$ decay products would produce a
multiplicity of particles.  Since the data set used to observe $\nu_\mu$
oscillation selects only single-particle events, only the resulting
disappearance of $\nu_\mu$ is thus observed.

An alternative scenario that can explain muon neutrino disappearance is
oscillation with a sterile neutrino ($\nu_s$), which undergoes neither
CC nor neutral current (NC) interactions.  A fourth flavor of neutrino
must be invoked to simultaneously explain apparent neutrino oscillations
in three widely separated regimes: atmospheric neutrinos; solar
neutrinos \citep{solarreview}; and the LSND experiment \citep{LSND}.
However, studies of the $Z^0$ at LEP \citep{z0} indicate that any such
new light neutrino flavor must not interact via the weak force, leading
to speculation about a sterile flavor.

Super--K has used two methods to distinguish between the $\nu_\tau$ and
$\nu_s$ hypotheses for explaining atmospheric $\nu_\mu$ disappearance
\citep{taust}.  The first method is to examine events likely to have
been caused by NC interactions.  While $\nu_\tau$ readily undergo such
interactions, $\nu_s$ would not, resulting in a relative suppression of
the NC signal.  The second method exploits so-called ``matter effects'',
which reduce the oscillation amplitude of high-energy
$\nu_\mu\leftrightarrow\nu_s$ in the presence of matter \citep{matter}.
While the combination of these two analyses disfavors
$\nu_\mu\leftrightarrow\nu_s$ compared with
$\nu_\mu\leftrightarrow\nu_\tau$, the observation of appearance of the
newly created $\nu_\tau$ would be a definitive answer.  Presented here
are an update to the NC and matter effect analyses and recent attempts to
resolve such $\nu_\tau$ appearance.

\section{Data Analysis}

Super--K is a 50~kt water Cherenkov detector employing 11,146
photomultiplier tubes (PMTs) to monitor an internal detector (ID) fiducial
volume of 22.5 kilotons. Entering and exiting charged particles are
identified by 1885 PMTs in an optically isolated outer volume (OD). Details
of the detector, calibrations, and data reduction can be found in
\citep{evidence,SKupmu}.  
Fully-contained (FC) events deposit all of their Cherenkov light in the
ID while partially-contained (PC) events have exiting tracks which
deposit some light in the OD.  The vertex position, number of Cherenkov
rings, ring directions, and momenta are reconstructed and the particle
types are identified as ``$e$-like'' or ``$\mu$-like'' for each
Cherenkov ring.  In the current 79.4 kiloton-year (1289 day) FC sample,
there are 3490 single-ring $e$-like events, 3346 single-ring $\mu$-like
events and 3477 multi-ring events.

Based on 1268 live days, this detector has also collected 1416 upward
through-going muon (UTM) events produced by atmospheric neutrino interactions
in the surrounding rock.  We required a minimum track length of 7~m in
the inner detector and an upward muon direction.  Downward-going
neutrino induced muons cannot be distinguished from the 3~Hz of cosmic
ray muons.  

\subsection{Observing $\nu_\mu$ disappearance}

The FC single-ring data sample was analyzed to establish the region in
($\sin^22\theta,\Delta m^2$) oscillation parameter space where $\nu_\mu$
disappearance is taking place.  This data sample is dominated by
quasi-elastic neutrino interactions, which allow the tagging of the
flavor of the parent neutrino via the flavor of the observed outgoing
lepton.  Data was binned by energy, outgoing lepton flavor, and zenith
angle of the lepton.  The distance the neutrino has traveled can be
inferred from the zenith angle, with those neutrinos coming from above
having traveled only tens of km, and those coming from below the whole
diameter of the earth.  To provide an expected signal, 70 live-years
equivalent of Monte-Carlo (MC) data was used.  As published by
\citet{evidence}, the number of of observed $\mu$-like events in bins
with low energies and long baselines is suppressed compared with the
non-oscillated MC expectations.

To evaluate the suitability of possible oscillation parameters, a MC
expectation was calculated using the oscillation probability appropriate
for both two-flavor $\nu_{\mu} \leftrightarrow \nu_{\tau}$ and
$\nu_{\mu} \leftrightarrow \nu_{s}$ oscillations.  A $\chi^{2}$
comparison of these MC predictions and the FC single-ring data yielded
allowed regions in ($\sin^22\theta,\Delta m^2$) parameter space for both
types of $\nu_\mu$ disappearance oscillations.  For the case of
$\nu_{\mu} \leftrightarrow \nu_{s}$, matter effects were included,
although at the $\sim 1$~GeV neutrino energies in this data sample the
matter effects are small and the two resulting allowed regions very
similar.  This analysis results in two sets of neutrino oscillation
parameters which are consistent with $\nu_\mu$ disappearance
oscillations for a $\nu_\tau$ oscillation partner or a $\nu_s$.  For
details of this fit see \citet{taust}.

\subsection{Neutral Current Sample}

\begin{figure}[t]
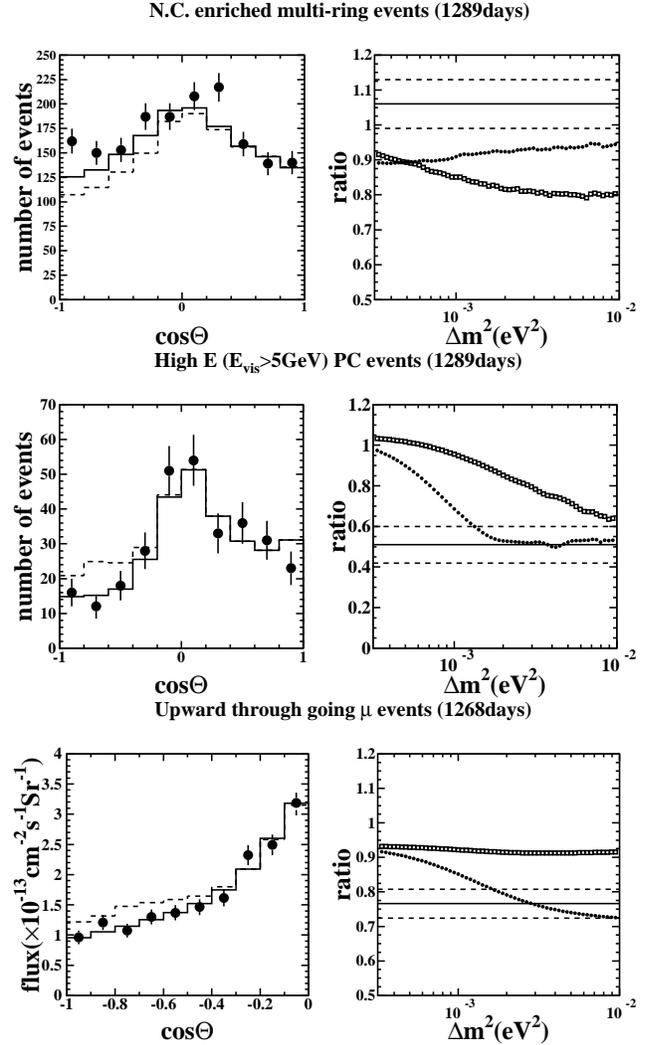

\vspace*{2.0mm} 
\includegraphics[width=8.3cm]{nczen.epsi}
\includegraphics[width=8.3cm]{pczen.epsi}
\includegraphics[width=8.3cm]{utmzen.epsi}
\caption{(a,c,e) Zenith angle distributions of atmospheric neutrino
  events satisfying cuts described in the text: (a) multi-ring sample,
  (c) partially contained sample, and (e) upward through-going muon
  sample.  The black dots indicate the data and statistical errors.  The
  solid line indicates the prediction for $\nu_\mu \leftrightarrow
  \nu_\tau$, and the dashed for $\nu_\mu \leftrightarrow \nu_s$, with
  (\dmsq,~$\sin^{2}2\theta$)=($3.2\times 10^{-3}$~eV$^{2}$,1).  The two
  predictions are independently normalized to the number of
  downward-going events for (a) and (c) and the number of horizontal
  events for (e).  (b,d,f) Expected value of the corresponding test
  ratio as a function of \dmsq. The solid horizontal lines indicate the
  measured value from the Super--K data with statistical
  uncertainty indicated by dashed lines. Black dots indicate the
  prediction for $\nu_\mu \leftrightarrow \nu_\tau$, and empty squares
  for $\nu_\mu \leftrightarrow \nu_s$, in both cases for maximal
  mixing.}
\label{fig:MR} 
\end{figure}

A set of cuts were devised to isolate a comparatively NC rich data set.
These events must be contained, neutrino-induced events with multiple
rings, the brightest of which must be identified as an electron.
Furthermore, to improve the angular correlation of the observed
particles to their parent neutrino,the total visible energy must be
greater than 400~MeV.  This results in a mean angle difference between
the parent neutrino and the reconstructed event direction of
$33^{\circ}$.  When tested on the MC data, these cuts produce a data
sample containing a 29\% fraction of NC events.  The FC single ring
sample discussed above contains only $\sim 6\%$ NC events.  Figure
\ref{fig:MR}(a) shows the zenith angle distribution of these events with
predictions from the MC.

Since $\nu_s$ do not interact via a NC channel, if the $\nu_\mu$ are
changing into $\nu_s$ given sufficient distance to do so then the
long-pathlength zenith angle bins coming from below will have
comparatively fewer NC events and the ratio of up-to-down events will
be reduced.  However, $\nu_\tau$ do interact via the NC channel.  Thus,
if the explanation of the mu-like event suppression observed in the FC
data is that $\nu_\mu$ are changing to $\nu_\tau$, then the NC event
rate coming up from below will remain unchanged, as will the up/down
ratio of NC events.

We define ``upward'' as a cosine of zenith angle less than $-$0.4 and
``downward'' as greater than +0.4.  There are 465 upward events and 438
downward events in the NC-enriched sample. Figure \ref{fig:MR}(b) shows
the $\Delta m^{2}$ dependence of the expected up/down ratio in the case
of full mixing ($\sin^2 2\theta = 1)$. For $\Delta m^2 = 3.2 \times
10^{-3}$~eV$^2$, the data are consistent with $\nu_{\mu} \leftrightarrow
\nu_{\tau}$, while the up/down ratio predicted by $\nu_{\mu}
\leftrightarrow \nu_{s}$ oscillations is 3.4$\sigma$ too low.

Using this up-to-down ratio helps to cancel some systematic
uncertainties, particularly the rather large uncertainties in the NC
cross-sections.  The total systematic uncertainty in the up/down ratio
of the data and MC is estimated to be $\pm2.9\%$ \citep{taust}.

\subsection{Partially Contained Sample}

To look for the presence of matter effects upon the oscillation
amplitude, we turn to a higher energy event sample as matter effects are
strongest at high energies.  The partially contained (PC) event sample
is created by neutrino interactions in the fiducial volume of water
which produce an outgoing particle with a high enough energy to escape
the inner detector.  PC events in Super--K are estimated to be
97\% pure $\nu_\mu$ CC and result from parent neutrinos with a mean of
10~GeV.  A further requirement of a visible energy of greater than 5~GeV
raises this mean neutrino energy to 20~GeV.  Figure \ref{fig:MR}(c)
shows the zenith angle distribution of these events with predictions
from MC, as before.  Again an up/down ratio is employed to cancel
systematic uncertainties using the same angular definition as for the
multi-ring sample.  There are 46 ``up'' events and 90 ``down'' events.
Figure \ref{fig:MR}(d) shows the $\Delta m^{2}$ dependence of the
expected up/down ratio in the case of full mixing.  Since matter effects
suppress oscillations to $\nu_s$, the up/down ratio would remain near
one in the case of $\nu_{\mu} \leftrightarrow \nu_{s}$ oscillations.
However, for $\Delta m^2 = 3.2 \times 10^{-3}$~eV$^2$, the observed
up/down ratio is consistent with $\nu_{\mu} \leftrightarrow \nu_{\tau}$
oscillations and 2.9$\sigma$ too low for $\nu_{\mu} \leftrightarrow
\nu_{s}$.  The total systematic uncertainty in the up/down ratio is
estimated to be $\pm4.1\%$.


\subsection{Upward Through-going Muon Sample}

An even higher energy sample of neutrinos is observed by Super--K and
produces upward through-going muon events.  These neutrinos interact in
the rock below Super--K and produce an outgoing muon with enough energy
to reach and completely traverse the detector.  The neutrinos which
produce such events have a peak energy of 100~GeV, maximizing any
possible matter effect suppression of oscillation amplitudes.  Due to
the down-going presence of cosmic ray muons, a ``vertical''
($\cos\theta<-0.4$) to ``horizontal'' ($-0.4<\cos\theta<-0$) ratio is
used to probe this sample.  Figure \ref{fig:MR}(e) shows the zenith
angle distribution of these events with predictions.  Figure
\ref{fig:MR}(f) shows the $\Delta m^{2}$ dependence of the expected
vertical/horizontal ratio in the case of full mixing. At the point of
$\Delta m^2 = 3.2 \times 10^{-3}$~eV$^2$, the data are consistent with
$\nu_{\mu} \leftrightarrow \nu_{\tau}$ oscillation, while the predicted
suppression of oscillations via the matter effect on $\nu_{\mu}
\leftrightarrow \nu_{s}$ oscillations is again not seen at a
significance of 2.9$\sigma$.  We estimated the total systematic
uncertainty in the horizontal/vertical flux ratio to be $\pm3.3\%$.

\subsection{Evaluating $\nu_{\mu}
  \leftrightarrow \nu_{\tau}$ versus $\nu_{\mu} \leftrightarrow
  \nu_{s}$}

All four data samples discussed (FC single ring, NC, PC, and UTM) are
independent.  No data found in one is present in another.  Thus, the
results of the four can be statistically combined to address the
question of $\nu_\tau$ versus $\nu_s$.  The FC data sample sets the
allowable range of (\sstt,\dms) that fit the observed $\nu_\mu$
disappearance oscillations for both $\nu_\tau$ and $\nu_s$.  Each pair
of oscillation parameters in those allowed regions is used to generated
an expected neutrino signal for each of the other three samples.  This
allows a comparison between the up/down or vertical/horizontal ratios
observed in each of the three samples and that expected for both
hypotheses.  The data samples being independent, the sum of the squares
of the three deviations between data and expectations should follow a
\chisq distribution with three degrees of freedom.  A (\sstt,\dms) pair
is disfavored at the 90(99)\% C.L. if the value of the sum is greater
than 6.3(11.3).  For the case of $\nu_\mu\leftrightarrow\nu_{s}$, all
parameters allowed at 99\% C.L. by the FC single-ring $\nu_s$ case are
disfavored by least the 99\% C.L.  when the other three data sets are
examined.  However, all oscillation hypotheses in the
$\nu_\mu\leftrightarrow\nu_\tau$ FC 90\% C.L. allowed region remain
allowed by the three additional data sets.

\subsection{Evaluating $\nu_\mu\leftrightarrow(\nu_\tau+\nu_s)$}

While the above analysis disfavors pure $\nu_\mu\leftrightarrow\nu_s$
oscillations, $\nu_s$ could still be present as a sub-dominant partner
in dominant $\nu_\mu\leftrightarrow\nu_\tau$ oscillations if $\nu_\tau$
and $\nu_s$ form a close mass doublet \citep{fogli}.  To test the extent
of such mixing which would produce an expectation which remains
compatible with the observed Super--K data, a global $\chi^2$ was formed
between all the atmospheric neutrino data present in Super--K (FC, PC,
UTM, multi-ring NC, multi-ring $\mu$-like, and upward-stopping muons)
and MC expectations including the appropriate level of $\tau$ appearance
and matter effects.  Systematic errors were treated in a similar fashion
to \citet{evidence}.

This global fit has 190 degrees of freedom.  A pure $\nu_\tau$
oscillation partner produces expectations which have a $\chi^2$
difference from the data of 175.0 at the best fit point.  Pure $\nu_s$
as the oscillation partner yields $\chi^2=204.8$ at its best fit, and is
disfavored by $>5\sigma$ compared to pure $\nu_\tau$.  If the fraction
of $\nu_s$ present in the oscillation sample ($\sin^2\xi$) is varied
between these extremes such that
$\nu_\mu\rightarrow(\cos\xi\nu_\tau+\sin\xi\nu_s)$, the maximum
fraction of $\nu_s$ allowed at the 90(99)\% C.L. by the complete set of
Super--K data is $\sin^2\xi= 25(35)$\%.  Thus, it is still possible for
$\nu_s$ to be present in the atmospheric neutrino oscillations as a
sub-dominant partner.

\section{$\tau$ Appearance Searches}

In contrast to looking for (and failing to find) hints of $\nu_s$, one
could directly detect the presence of the newly created $\nu_\tau$ by
observing the $\tau$ appearance associated with a $\nu_\tau$
interaction.  This is difficult, because the $\tau$ itself has such a
short lifetime that one must look for its decay products and then deduce
that a $\tau$ was their parent.  Such events are both rather complicated
and comparatively rare, as only a small component of the atmospheric
neutrino spectrum exceeds the 3.4~GeV $\tau$ production threshold.
Around 75 such interactions are expected to be in the existing data set.
Additionally, neutrino interactions which produce a number of pions or
other mesons can mimic a $\tau$ decay within the finite resolution of
the Super--K detector.  However, three separate analyses have looked for
$\tau$ appearance in Super--K, concentrating on hadronic decay modes.
All three were developed using the existing atmospheric neutrino MC as
background with a $\tau$ signal injected.
To evaluate their expected sensitivity, the resulting analyses were also
applied to a MC atmospheric neutrino sample with $\tau$ included at the
level of maximal mixing and $\Delta m^2 = 3.0\times10^{-3}$, the
remaining range of allowable oscillation parameters being considered
here as a systematic error.

The first analysis makes use of the standard Super--K event
reconstruction tools, which use likelihood calculations to compute to
what degree the observed light pattern matches the expected distribution
for a given set of particles.  To isolate $\tau$ events, cuts are
imposed on the visible energy, number of rings, number of decay
electrons, the fraction of energy in the brightest ring, the distance
from the first ring to the decay electron, the max muon momentum, the
transverse momentum, and the particle ID of the brightest ring.

The second analysis makes use of a neural-net algorithm, using variables
obtained from the likelihood-based tools as input: number of rings
found; number of seeds used as the input to the ring counting algorithm;
number of decay electrons; particle ID of most energetic ring; number of
muons; and visible energy.  The net was trained and tested on $\tau$ MC
data and asked to classify data events as $\tau$ or not.  

The third analysis makes use of the fact that a $\tau$ produced near
threshold will decay symmetrically, while a non-$\tau$ pion-production
event will have remaining momentum and skew the resulting particles in
one direction.  Thus, energy flow and jet-finding calculations are made
in addition to initial cuts from the standard likelihood analysis.

\begin{table}[htbp]
  \begin{center}
    \begin{tabular}{|c|c|c|c|c|}\hline
      Analysis &$\epsilon_\tau$ & fit \#$\tau$& $\sigma_{exp}$&$\sigma$\\\hline
      Likelihood & 43\% & $66\pm 41_{stat}{^{+25}_{-18}}$&2.0&1.5\\
      Neural Net & 51\% & $92\pm 35_{stat}{^{+17}_{-23}}$&2.0&2.2\\
      Energy Flow & 32\% & $79^{+44}_{-40}(stat+sys)$&1.9&1.8\\\hline
    \end{tabular}
    \caption{$\tau$-appearance analyses, their efficiencies at saving
      $\tau$ events, the number of $\tau$ they observed, and the
      expected and observed significances of the signals.}
    \label{tab:tautab}
  \end{center}
\end{table}

After development and testing using the MC data as inputs, the analyses
were then applied to the atmospheric neutrino data set.  The events were
binned into five bins in $\cos\Theta$, since if $\nu_\tau$ are being
created via oscillations they will be coming upward (at long
pathlengths).  A comparison between the data, the MC expectations with
no $\tau$ appearance included, and with an expectation including $\tau$
appearance was made.  
The significances of the potential $\tau$ excess are around $2\sigma$
for all three analyses.  A fit to the amount of $\tau$ appearance which
best describes the data yields the number of observed $\tau$ shown in
Tab.~\ref{tab:tautab}.  These three analyses are highly correlated, so
one cannot combine their results in any statistically meaningful way.
While all three are consistent with the presence of $\tau$
appearance, none are statistically significant.


\section{Summary and Conclusion}

Three independent data samples are presented that discriminate
between the oscillations to either $\nu_\tau$ or $\nu_s$ in the region
of (\sstt,\dms) parameter space favored by the majority of the Super--K
data.  While two-flavor $\nu_\mu\leftrightarrow\nu_s$
$\nu_\mu\leftrightarrow\nu_\tau$ oscillations both fit the low energy
charged current data, the pure $\nu_s$ hypothesis does not fit the
higher energy and neutral current samples.  
Pure $\nu_{\mu} \leftrightarrow \nu_{\tau}$ neutrino oscillations do fit
all of the data samples presented, although a simultaneous fit to all
available Super--K atmospheric neutrino data allows up to a 25\% (90\%
C.L.) contribution from $\nu_s$ as a sub-dominant oscillation partner.
Additionally, searches for the $\tau$ appearance associated with
$\nu_{\mu} \leftrightarrow \nu_{\tau}$ oscillations result in signals
which are consistent with the presence of $\tau$ appearance in the
Super--K detector.

\begin{acknowledgements}
  We gratefully acknowledge the cooperation of the Kamioka Mining and
  Smelting Company. The Super--K experiment was built and has been
  operated with funding from the Japanese Ministry of Education,
  Science, Sports and Culture, and the United States Department of
  Energy. We gratefully acknowledge individual support by the National
  Science Foundation and Research Corporation's Cottrell College Science
  Award.
\end{acknowledgements}


\begin{thebibliography}{99}
\bibitem[Y.~Fukuda {\it et al.}(1998)]{evidence}
  Y.~Fukuda {\it et al.},
  Phys. Rev. Lett. {\bf 81}, 1562 (1998).
  
\bibitem[Y.~Fukuda {\it et al.}(1999)]{SKupmu} Y.~Fukuda, {\it et al.},
  Phys. Rev. Lett. {\bf 82}, 2644 (1999); Phys. Lett. B {\bf 467}, 185
  (1999).
  
\bibitem[Bahcall {\it et al.}(1998)]{solarreview} 
  J.N.~Bahcall, P. I. Krastev, \&
  A.Yu.~Smirnov, Phys. Rev. D {\bf 58}, 096016 (1998).

\bibitem[C.~Athanassopoulos {\it et al.}(1998)]{LSND}
  C.~Athanassopoulos {\it et al.},
  Phy. Rev. Lett. {\bf 81}, 1774 (1998).


\bibitem[Decamp {\it et al.}(1992)]{z0}
D.~Decamp {\it et al.}, Z. Phys. C {\bf 35}, 1 (1992);
P.~Abreu {\it et al.}, Nucl. Phys. B {\bf 367}, 511 (1992);
B.~Adeva {\it et al.}, Z. Phys. C {\bf 51}, 179 (1991);
G.~Alexander {\it et al.}, Z. Phys. C {\bf 52}, 175 (1992).

\bibitem[Y.~Fukuda {\it et al.}(2000)]{taust}
  Y.~Fukuda {\it et al.},
  Phys. Rev. Lett. {\bf 85}, 3999 (2000).



\bibitem[Liu {\it et al.}(1998)]{matter}
Q.Y.~Liu, S.P.~Mikheyev, \& A.Yu.~Smirnov, 
Phys. Lett. B {\bf 440}, 319 (1998) and many other references cited therein.

\bibitem[Fogli {\it et al.}(2001)]{fogli}G.L.~Fogli, E.~Lisi,
  A.~Marrone, Phys.Rev. D {\bf 63}, 053008 (2001).

  



\end{thebibliography}
\end{document}